\documentclass[nature,superscriptaddress,dvipsnames,twocolumn]{revtex4-1}

\usepackage{fancyhdr} 
\usepackage{layout}
\usepackage{color}
\usepackage{amsmath}
\usepackage{tikz}
\usepackage{xcolor}
\usetikzlibrary{arrows}

\usetikzlibrary[topaths]

\usepackage{siunitx}
\usepackage{eurosym}
\usepackage{amsfonts}
\usepackage{amsmath}
\usepackage{amsthm}
\usepackage{amscd}
\usepackage{amssymb}
\usepackage{amsxtra}
\usepackage{bm}
\usepackage{bbm}         
\usepackage{exscale}
\usepackage{graphics,color}
\usepackage{latexsym}
\usepackage{enumerate}
\usepackage{dcolumn}      
\usepackage{graphics}
\usepackage{epstopdf}
\usepackage[dvips]{epsfig}
\usepackage{graphicx}
\usepackage{hyperref}
\usepackage{tabularx}
\usepackage[authormarkup=none]{changes}
\usepackage[caption=false]{subfig}
\usepackage{lineno}
\usepackage[utf8]{inputenc} 
\usepackage{eurosym}

\usepackage{setspace}

\hypersetup{
    colorlinks,%
    citecolor=black,%
    filecolor=black,%
    linkcolor=black,%
    urlcolor=black
}

\newcommand{\ket}[1]{\ensuremath{|#1\rangle}}

\newcommand{\beq}{\begin{equation}}
\newcommand{\eeq}{\end{equation}}

\newcommand{\bdf}{\begin{defn}}
\newcommand{\edf}{\end{defn}}

\definechangesauthor[name={Rob}, color={orange}]{rob}

\newtheorem{defn}{Definition}

\newcount\mycount

\begin{document}



\title{\textbf{Progress in satellite quantum key distribution}}
\author{Robert Bedington}
\thanks{Authors contributed equally to this work}
\affiliation{Centre for Quantum Technologies, National University of Singapore}
\thanks{S15-02-10, 3 Science Drive 2, Singapore 117543}

\email{r.bedington@nus.edu.sg}
\thanks{tel:+65 65166758}

\author{Juan Miguel Arrazola}
\thanks{Authors contributed equally to this work}
\affiliation{Centre for Quantum Technologies, National University of Singapore}
\author{Alexander Ling}
\thanks{Authors contributed equally to this work}
\affiliation{Centre for Quantum Technologies, National University of Singapore}
\affiliation{Department of Physics, National University of Singapore}

\date{\today}

\begin{abstract}

Quantum key distribution (QKD) is a family of protocols for growing a private encryption key between two parties. Despite much progress, all ground-based QKD approaches have a distance limit due to atmospheric losses or in-fibre attenuation. These limitations make purely ground-based systems impractical for a global distribution network. However, the range of communication may be extended by employing satellites equipped with high-quality optical links. This manuscript summarizes research and development which is beginning to enable QKD with satellites. It includes a discussion of protocols, infrastructure, and the technical challenges involved with implementing such systems, as well as a top level summary of on-going satellite QKD initiatives around the world.

\end{abstract}

\begin{titlepage}
\maketitle
\clearpage

\end{titlepage}

\section{Introduction}
Quantum key distribution (QKD) is a relatively new cryptographic primitive for establishing a private encryption key between two parties. The concept has rapidly matured into a commercial technology since the first proposal emerged in 1984~\cite{BB84}, largely because of a very attractive proposition: the security of QKD is not based on the computational hardness of solving mathematical problems, but on physical processes that are not vulnerable to powerful computers. As an example, in public-key encryption, the public keys and message transcripts can be stored and subjected to cryptanalysis at any time, and while they might be secure now, they might not be secure against newer, more powerful computers at some point in the future. In contrast, if the key generated from a QKD protocol is secure today, it will remain secure against advances in computing power. This property, known as `forward security', makes it an ideal solution to ensure the secrecy of sensitive data that must be kept confidential for long periods of time. Furthermore, as QKD is primarily an optical technology, it has the ability to automate the delivery of encryption keys between any two points that share an optical link, which is advantageous given the growth of optical communication networks. In particular, QKD has the potential to replace or augment existing trusted-courier systems for secure transmission of encryption keys.

There are a variety of optical techniques for implementing QKD. The most common class of solutions encode each bit of private information onto discrete degrees of freedom of optical signals, and hence is termed collectively as discrete-variable QKD (DV-QKD). An alternative approach employs coherent communication techniques to encode the private information, and is known as continuous-variable QKD (CV-QKD). Both approaches have seen dedicated engineering that has led to increased key generation rates and improved compatibility with current communications infrastructure \cite{dixon2015high,choi2014field,frohlich2015quantum}. However, both approaches face a similar obstacle when attempting to implement wide-scale deployment of QKD: physical communication channels introduce transmission losses that increase exponentially with distance, greatly limiting the secure key rates that can be achieved over long ranges. 

For any pure-loss channel with transmittance $\eta$, it has been shown that the secure key rate per mode of any QKD protocol scales linearly with $\eta$ for small $\eta$ \cite{Takeoka2014,Pirandola2015a}. This places a fundamental limit to the maximum distance attainable by QKD protocols relying on direct transmission. To put this into perspective, consider that a conventional telecommunications fibre with an attenuation of \SI{0.18}{dB/km} \cite{smf28} stretching over \SI{1000}{km} has a theoretical transmittance of precisely $\eta=10^{-18}$ -- worse in real-world deployments -- making the resulting key rates forbiddingly small. The problem remains even if using state-of-the-art fibres which can reach attenuations of  \SI{0.142}{dB/km} \cite{Tamura2017}. Experiments have been reported where QKD was performed for distances of up to \SI{404}{km} \cite{Yin2016}, but the resulting key rates remain several orders of magnitude smaller than the requirements for practicable deployment.  Sufficiently large key rates can only be obtained over metropolitan-scale networks where the range is within \SI{100}{km}.

In principle, it is possible to extend the range of QKD by using quantum repeaters. Some proposals for quantum repeaters rely on heralded entanglement generation and purification \cite{sangouard2011,duan2001}, while other recent architectures employ logical qubits capable of quantum error-correction to overcome both operational errors \cite{fowler2010,jiang2009} and errors due to loss \cite{muralidharan2014,munro2012}. The difficulty of constructing such systems, however, is comparable to the challenges faced in building universal quantum computers, with the required technology unlikely to be available in the near future. 

An alternative, which is viable with the present level of technology, is to use satellites as intermediate trusted nodes for communication between locations on the ground. By placing a satellite above the Earth's atmosphere, direct links can be established between ground stations and the satellite, thus enabling communication between distant points on the planet. Atmospheric attenuation in free space is less significant than in fibre, where for instance values of $0.07$ dB/km can be achieved at \SI{2400}{m} above sea-level \cite{schmitt2007}, with higher attenuations at lower altitudes and negligible attenuation in the vacuum above the Earth's atmosphere. Dominant sources of loss in this case occur due to beam diffraction and the limited size of telescopes at the receiver. However, the overall effect is a greatly reduced amount of loss compared to ground-level transmission, making satellite QKD a promising route to enable secure key generation across global distances. Consequently, a number of efforts are under way to raise the technology readiness level of satellite QKD. These projects range from truck-based tests of pointing and tracking mechanisms \cite{bourgoin2015free}, to in-orbit testing of quantum light sources \cite{Tang2016}, to full QKD demonstrations using orbiting satellites \cite{Pan2014}. Together, these efforts will enable global QKD services as well as advanced fundamental experiments \cite{Rideout2012}.

In this review, we give an overview of the advances, challenges, and future directions of satellite-based QKD. We begin by discussing the basic concepts for performing QKD with satellites, regarding both protocols and infrastructure. We continue by discussing the technical challenges and progress towards realization in more detail. We conclude by giving a summary of existing quantum satellite initiatives around the world.

\section{QKD Protocols}
In this section, we give a brief overview of QKD as well as the most commonly deployed protocols, which can be skipped by experts in the field. An in-depth review of QKD can be found in Ref. \cite{scarani2009security} while a discussion of the frontiers of research in practical QKD appears in Ref. \cite{Diamanti2016}.  

QKD is a scheme for enabling two parties, commonly referred to as Alice and Bob, to derive a private and symmetric encryption key. As discussed in the previous section, QKD protocols can be divided into two main classes: DV-QKD or CV-QKD. In CV-QKD, information is encoded in the quadratures of randomly selected coherent states, which are then measured using either homodyne or heterodyne detection \cite{grosshans2003,diamanti2015}. From an implementation perspective, the optical tools are already very mature and widely used in conventional coherent communication schemes.  However, significant challenges remain, such as in the need to decrease the classical communication between the two parties when establishing the final, secret key \cite{Diamanti2016,jouguet2013experimental,diamanti2015}. Most satellite QKD projects have chosen to implement discrete-variable schemes, and this paper will concentrate on DV-QKD concepts and technical solutions.

DV-QKD systems can be subdivided into prepare-and-measure or entanglement-based protocols. In a generic prepare-and-measure protocol \cite{BB84,B92,scarani2004quantum}, Alice encodes each classical bit into an individual optical signal before transmitting it to Bob, who performs a prescribed set of measurements on each of the incoming signals in order to retrieve the classical data encoded in their states. 

In the BB84 protocol \cite{BB84}, which is the most widely deployed prepare-and-measure protocol, each classical bit is encoded into the polarization of an individual photon. Alice prepares the photon states by randomly choosing between the horizontal/vertical $Z$ basis $\{\ket{H},\ket{V}\}$ and the +\ang{45}/-\ang{45} $X$ basis $\{\frac{1}{\sqrt{2}}(\ket{H}+\ket{V}),\frac{1}{\sqrt{2}}(\ket{H}-\ket{V})\}$, where $\ket{H},\ket{V}$ are single-photon states of horizontal and vertical polarization. The states in each basis are assigned bit values of 0 and 1 respectively, from which the encryption key can be established.

Once Alice has selected the basis, she randomly selects one of the two states and sends it to Bob. Upon receiving a signal from Alice, Bob randomly selects one of the two bases and performs a corresponding projective measurement, recording his outcome as a classical bit. After many signals have been sent, Alice and Bob publicly announce the basis they have chosen for each signal and they discard all cases when they chose a different basis. Finally, they select a random subset of their data to estimate their relative errors. If the error rate is sufficiently low, they apply error-correction and privacy amplification to their data to reach a final shared secret key. An important consideration is that the transmitter, Alice, must have access to an active source of good random numbers in order to make both the basis choice and state choice. The recipient, Bob, needs also to be able to make a random choice on the measurement set, but this can be done passively, using beamsplitters \cite{Marcikic2006}. 

Due to a lack of robust, true single-photon sources, Alice's states are usually weak coherent states produced from laser sources, where each pulse has a finite probability of containing more than a single photon. To be more resilient against photon-number splitting attacks, an important innovation has been the creation of `decoy-states' \cite{lo2005decoy,hwang2003}, where Alice also chooses randomly between two intensities of her coherent state signals, which she reveals publicly to Bob after the quantum communication has been completed. The use of an additional degree-of-freedom, such as the intensity leads to improved tolerance to losses compared to regular BB84. Notably, these protocols can be currently deployed with commercial off-the-shelf components.

Entanglement-based protocols \cite{BBM92,ekert1991} differ from prepare-and-measure systems by removing the need for an active choice when encoding states into the photons. Instead, both parties are recipients who share a source of maximally-entangled photon pairs. Typical implementations utilize photon pairs entangled in the polarization degree of freedom, and the photon pair is split such that one photon is transmitted to Alice, while its twin is transmitted to Bob. Both parties make independent measurement choices on the photons, and  decide to measure them in either the $X$ or $Z$ basis. In the most popular entanglement-based protocol known as BBM92 \cite{BBM92}, Alice and Bob perform parameter estimation, error correction, and privacy amplification in the same manner as in the BB84 protocol. This protocol has the advantage that no active random number generators are required for preparing the source, and that the measurement devices for Alice and Bob are identical. 

In an alternative entanglement-based approach, the E91 protocol \cite{ekert1991}, the measurement scheme must result in a statement on whether the photon-pair correlations between Alice and Bob violate the Bell inequality. This protocol is the precursor of device-independent QKD schemes proposed much more recently \cite{Acin2007}. From a practical perspective, the measurement scheme in E91 is less efficient in its use of photon pairs as the Bell inequality test requires more polarization settings to be monitored \cite{Ling2008b}.


\begin{figure}[h!]
\includegraphics[width=\columnwidth]{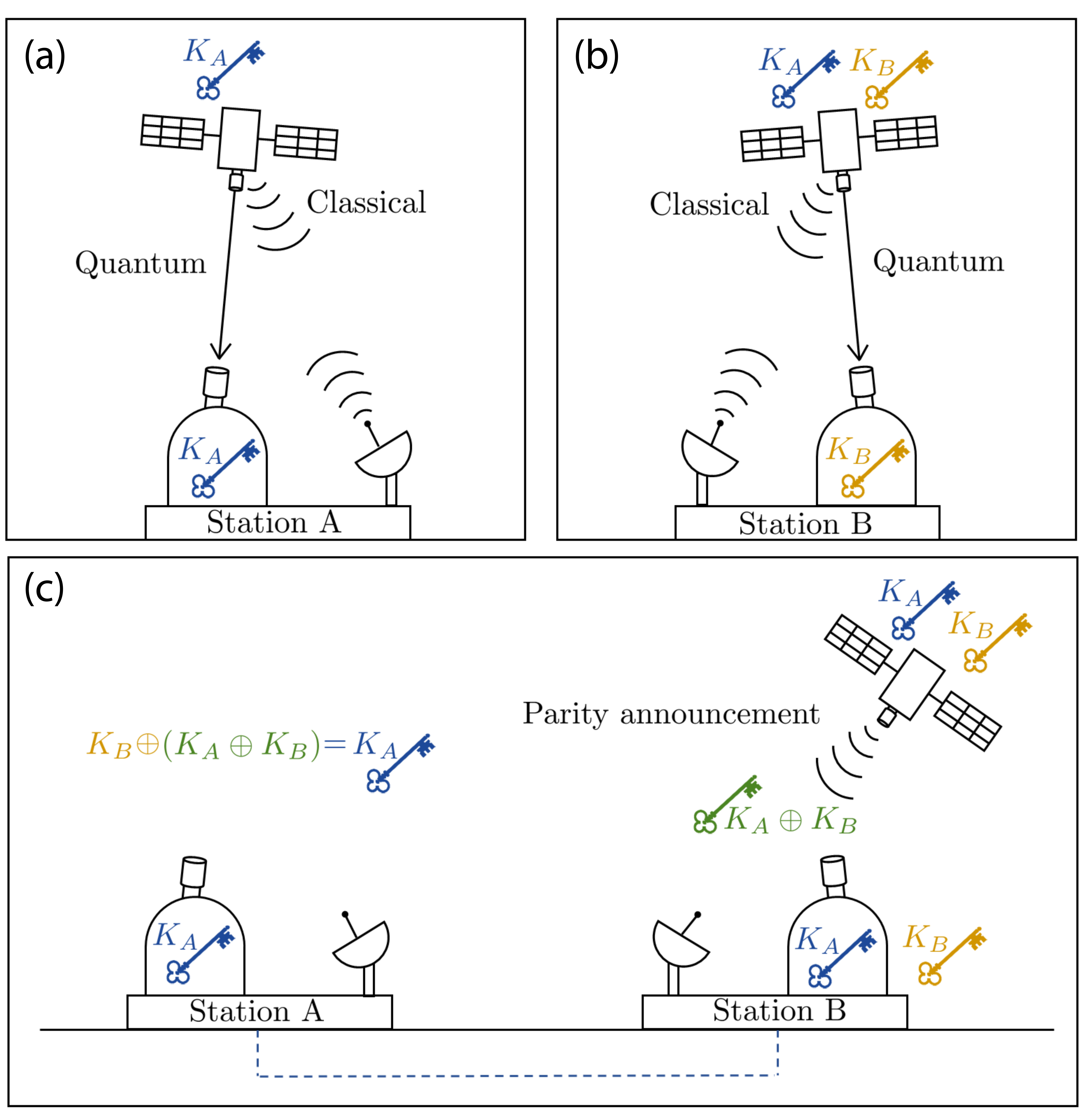}
\caption{\label{fig:protocol}Illustration of the most common satellite QKD scheme: the flying trusted-node. In step (a), the satellite establishes a shared secret key $K_A$ with station A by running a QKD protocol, which requires both classical and quantum communication. This step is repeated in (b) to establish a shared secret key $K_B$, this time with station B located further away. At the end of these steps, the satellite holds both keys, while each station knows only their own. Finally, in step (c), the satellite publicly announces the parity of both keys $K_A\oplus K_B$. This allows station B to determine key $K_A$, which can then be used to encrypt private communications to A and vice versa.}
\end{figure}

\section{Concepts for satellite QKD}
\label{sec:concepts}
In this section, we provide a high-level description of some different conceivable approaches for performing satellite QKD, which can be classified depending on the types of communication links that can be established, as well as the orbit of the satellite.

From a general perspective, most projects envision the satellite as a flying trusted-node (see Fig~\ref{fig:protocol}). In this scenario, the satellite carries out QKD operations with distinct ground stations to establish independent secret keys with each of them. The satellite holds all keys, while the stations only have access to their own keys. To enable any pair of stations -- for example station A and station B --  to share a common key, the satellite combines their respective keys $K_A$ and $K_B$ and broadcasts their bit-wise parity $K_A\oplus K_B$. Using this announcement, the stations can retrieve each other's keys because $K_A\oplus(K_A\oplus K_B)=K_B$ and $K_B\oplus(K_A\oplus K_B)=K_A$. Since the original keys are independent secret strings, their bit-wise parity is just a uniformly random string, so the parity announcement does not reveal any useful information to potential eavesdroppers. However, since the satellite holds all keys, access to the data obtained by the satellite would give an adversary complete knowledge of the key. Therefore, in this setting, the satellite must be trusted.

Trusted-node topologies are being built for fibre networks \cite{shanghai}, but ground-based nodes are fixed locations that can be subject to constant surveillance and probes. Side-channel attacks on QKD hardware that require access to the optical link \cite{Gerhardt2011} can face significant challenges in satellite scenarios where the optical link is moving relatively quickly. Most side-channel attacks are aimed at retrieving the key, and are distinct from denial-of-service attacks where the goal is to disable the satellite receiver, e.g. by aiming a sufficiently intense beam at the satellite optical transceiver. 
Denial-of-service attacks may be less of an issue if the satellite is only a transmitter. Additional research is needed however, to better understand potential vulnerabilities in satellite QKD.

\begin{figure*}[t!]
\includegraphics[width=0.8\textwidth]{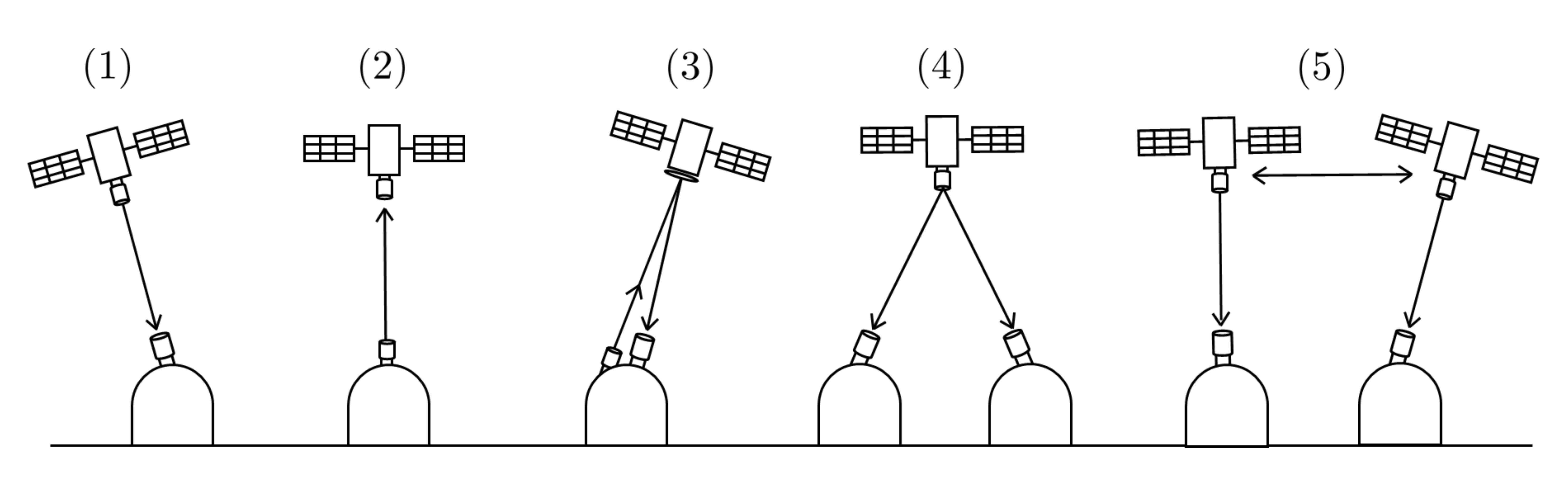}
\caption{\label{fig:scenarios} Illustration of different platforms for performing satellite QKD. Scenarios (1) and (2) depict a downlink and an uplink, respectively, while in scenario (3) a downlink is simulated by using a retro-reflector on board the satellite. In (4) pairs of entangled photons are being transmitted to Earth so that two ground stations can share entangled states. Finally,  scenario (5) illustrates how inter-satellite links can allow more complex satellite QKD networks.}
\end{figure*}

Quantum communication links with a satellite can be classified either as uplinks -- where the ground station sends signals to a receiver in space -- or downlinks, where the satellite sends the signals to the ground. Correspondingly, there are several possible configurations for performing QKD with satellites depending on the types of links that are used \cite{nordholt2002present,Aspelmeyer2003,rarity2004quantum,PerdiguesArmengol2008,Rideout2012}. These are illustrated in Fig.~\ref{fig:scenarios}. 


There are advantages and disadvantages associated with each configuration, but the more commonly recommended scenario for operational QKD -- and the only one that has thus far been demonstrated \cite{Sheng2017} -- is to use downlinks. This is because downlinks always have lower losses for any ground-satellite segment. This arises because atmospheric properties such as turbulence cause the optical beam to wander, which translates into a less accurate ground transmitter compared to a space-based transmitter. This is discussed in more detail in section~\ref{sec:opt}.

The main advantage of using an uplink is that it is not necessary to locate a (potentially complex) quantum light source in space, but only to place a receiver on board the satellites \cite{Jennewein2014d, Ren2017}. 
It also makes attacks that target receivers significantly more difficult  \cite{Makarov2016}.

It is also possible to achieve a downlink by using retro-reflectors on the satellite, which modulate signals sent from the ground as they bounce off back to a receiver also on the ground \cite{Vallone2015}. The challenge here is to develop fast modulating retro-reflectors, and to develop countermeasures that prevent an eavesdropper from sampling the state of the retro-reflector while QKD is being carried out.

Using downlinks also gives rise to the possibility of option 4 in Fig.~\ref{fig:scenarios}. A source of entangled photon pairs can be located in the satellite, and all of the photon pairs transmitted to ground---one photon in each pair to one ground station, and the other in each pair to a second ground station \cite{PerdiguesArmengol2008, Pan2014}. This configuration allows the realization of entanglement-based QKD directly between the ground stations, without using the satellite as a trusted node. 
The Chinese Micius satellite \cite{Yin2017} has successfully demonstrated entanglement swapping over \SI{1200}{km} to two mountain-top observatories using this, option 4, configuration. The \SIrange{64}{70}{dB} losses they experience are comparable with those predicted in studies \cite{Bourgoin2012}  but make practicable QKD challenging.

In the Micius demonstration a Bell inequality test between the two ground stations proves that the photons are entangled. Since the photons are entangled, no other party, including the satellite, has made a measurement on them previously. This scenario only works when both ground stations simultaneously have line-of-sight to the satellite. Also at wider separations slant angles would be low, so the optical link would have to traverse longer distances, and thus more atmosphere, than a direct, overhead pass. The losses are so high because the losses in both arms are effectively combined since only photon pairs that arrive at both stations can be used.

It has also been suggested that sources of entangled photons in space could be employed as quantum repeater stations, enabling entanglement swapping among more distant locations on the ground \cite{boone2015}. As more quantum ground stations come online more complex network architectures will become necessary \cite{Sheng2016} and will likely include inter-satellite quantum links \cite{Pfennigbauer2003}. 

The characteristics of the communication links are themselves dependent on the specific orbit of the satellite \cite{rarity2004quantum}. Orbital altitudes are classified into three main categories: Low Earth orbit (LEO), Medium Earth orbit (MEO), and Geostationary orbit (GEO) \cite{wertz1991space}. LEO are situated between 160 to \SI{3000}{km} in altitude (usually below \SI{900}{km}), GEO has an altitude of precisely \SI{35786}{km}, while MEO correspond to all orbits between LEO and GEO.  For satellite QKD applications, LEO and GEO are the two most suitable options. Although most current programmes opt for LEO, future projects might seek higher altitudes \cite{Sheng2016}. A satellite in LEO benefits from proximity to the surface which significantly reduces losses due to beam diffraction. The drawbacks in this case are the high speed of the satellite relative to surface of the Earth, which makes it challenging to achieve accurate pointing during signal transmission, as well as the fact that QKD can only be performed during the limited flyover time of the satellite. On the other hand, the situation is essentially reversed in GEO; the satellite is at rest relative to the ground, but it is located at a much higher altitude, where QKD can in principle be run continuously, but at the expense of much higher losses. 

Polar LEO will pass ground stations at the poles with every orbit and equatorial LEO will pass ground stations on the equator with every orbit, but other orbital inclinations and ground station locations will have less regular passes. The time of day of the flyovers will also vary unless the orbit is specially chosen to be sun synchronous. The Micius satellite for example is in a \SI{500}{km} sun synchronous orbit so that it passes over the Xinglong ground station for 5 minutes every night at around 12:50am local time \cite{Sheng2017}. In comparison, a night-time-operating QKD satellite in the ISS (International Space Station) orbit -- which is partway between a polar and an equatorial orbit -- would have about 150 usable night time passes over a  ground station at \ang{40} latitude during a period of 12 months \cite{Oi2017}.

It should be noted that GEO orbits place the satellites above the equator, so the satellite is closer to the horizon for locations that approach the poles, disappearing below the horizon at \ang{81} latitude. In these cases, the optical link must traverse a much larger amount of atmosphere and will suffer additional loss. Satellite QKD for such locations might adopt less conventional orbit choices, such as  the Molniya highly elliptical orbits (HEO) \cite{Kidder1990} used to provide polar regions with near-constant satellite coverage.

\section{Technical Realization}
Satellite QKD proposals typically employ polarization encoding of individual photons, as this is well suited to free-space communication. 
When using polarization encoding, the photons are obtained either from a weak coherent pulse (WCP) source or from polarization-entangled photon-pair sources.
A minority of proposals investigate options for time-bin entanglement \cite{Vallone2016}, or QKD with orbital angular momenta of photons \cite{Mafu2013}, but these approaches are less mature. Their appeal lies in the possibility of generating hyper-entangled states, where multiple quantum states can be encoded into a single photon. 

Development towards the realization of polarization encoded, satellite-based QKD can be traced to the early demonstrations of free-space QKD using both weak coherent pulses and entangled photons. By 1998, Buttler et al. had produced \SI{1}{\km} outdoor QKD links for night-time operations \cite{Buttler1998}, later extending that to \SI{1.6 }{\km} with daytime operations \cite{Buttler2000}. Other research groups went on to achieve longer links, from \SI{10}{\km} \cite{Hughes2002a} to \SI{23.4}{\km} \cite{Kurtsiefer2002} to \SI{144}{\km} \cite{schmitt2007}. In 2016 the Micius satellite began performing QKD between space and ground which, at its maximum, is spanning a distance of \SI{1200}{km} \cite{Sheng2017}.

Space-based instruments represent serious design challenges, as size, weight and power are limited on spacecraft.
Robustness is also important as the satellite launch process is not gentle, yet precise optical alignments must be maintained. 
Finally, the space environment requires special considerations to allow for the satellite to survive the vacuum, microgravity, radiation and thermal environments it experiences in orbit \cite{Tan2015}. 

In the subsections below we review some technical details and challenges that have been published in this field, from the photon sources, to the optical links, to the quantum receivers and communications overheads.

\subsection{Photon sources}
\label{sec:sources}
\begin{figure}[h!]
\centering
\includegraphics[width=\columnwidth]{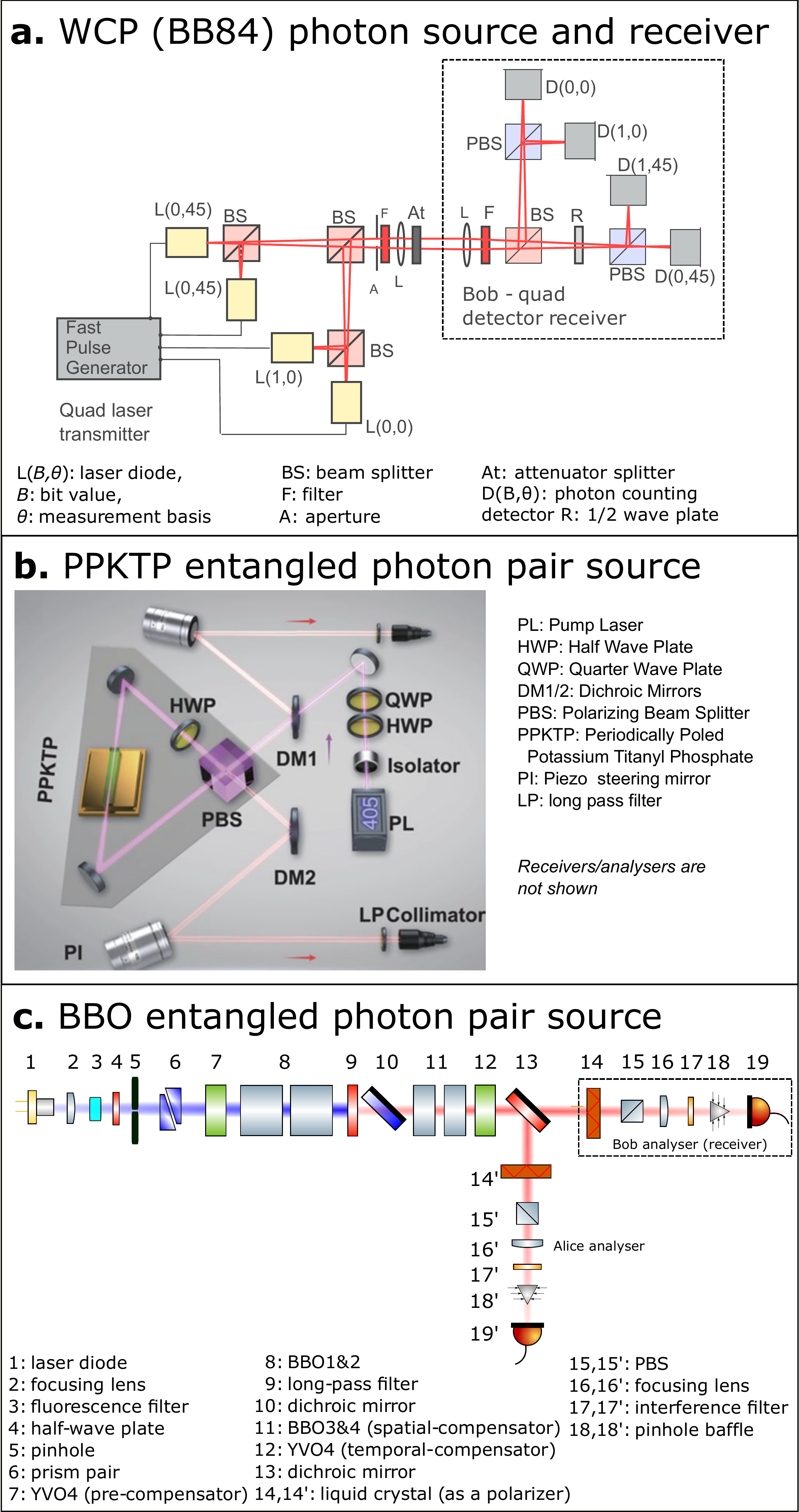}
\caption{\label{fig:sources} Examples of optical layouts for sources used in polarization encoded DV-QKD designed for satellite QKD.
(a) -- a weak coherent pulse source for BB84 QKD (adapted from \cite{rarity2002ground}).
(b) -- a polarization-entangled photon-pair source using PPKTP in a Sagnac loop arrangement (adapted from \cite{Yin2017}) 
(c) -- a polarization-entangled photon-pair source based on BBO with Bell's violation analysers (adapted from \cite{Durak}), for BBM92 QKD these analysers would most likely be replaced with quad detector receivers as per (a).}
\end{figure}

Significant progress has been made in the design of weak coherent pulse sources used to generate photons in most BB84 schemes \cite{Weier2006, Jofre2011a} (e.g. Fig. \ref{fig:sources}a). One of the bottlenecks towards high-speed BB84 schemes was the need for active polarization manipulation, which is both power-intensive and slow. This limitation was overcome in a design which used four laser diodes inside a single transmitter \cite{Weier2006}. By utilizing the high degree of polarization intrinsic to the diodes, this enabled each diode to be identified with a unique polarization state. A drawback of this approach is that the spectra of the diodes are not always identical and provide a possible side-channel for eavesdroppers \cite{Nauerth2009}. This side-channel was closed in a recent development where a single laser diode was coupled to four (direct-write) waveguides, each of which was capable of a fixed amount of polarization rotation \cite{Vest2015}. The waveguides were then recombined to result in a single-mode output with four possible polarization states. An added advantage of this design is its small form-factor, enabling it to be considered as a transmitter for both ground or satellite segments.

The Micius satellite performs space-to-ground BB84 QKD using \SI{850}{nm} photons from eight fibre-based laser diodes -- four for signal and four for decoy states -- pulsed at \SI{100}{MHz} \cite{Sheng2017}. An alternative design could use variable strength lasers, which might additionally be able to function as a laser beacon (see section~\ref{sec:pat}) \cite{Oi2017}.

Retroreflector schemes use ground-based light sources, which can be quite powerful -- not themselves WCP -- because it is only the reflected pulses from the retroreflector that need be in the weak coherent state. 
For synchronization purposes, such QKD photons could be combined with the pulse train from a satellite laser ranging system \cite{Vallone2015}. 
To use these setups to perform BB84 QKD the photons leaving the source would all have the same polarization, which would be modified by the satellite upon reflection by a polarization-modulating retroreflector. 
Significant development work is likely to be necessary to produce a retroreflector that could be modulated at sufficiently high rates with sufficiently high contrast ratios. 
There would also need to be various compensation devices to overcome the effects of the satellite motion, and as with many QKD systems, irradiance measurements would need to be taken to look out for eavesdroppers \cite{Gerhardt2011}, i.e.~to check that no eavesdropper laser was also interrogating the retroreflector \cite{Gisin2006}.

The QKD schemes that utilize entanglement almost all rely on polarization entangled photon-pair sources based on bulk-crystal, collinear, spontaneous parametric downconversion (SPDC), either periodically-poled potassium titanyl phospate (PPKTP) or single-domain crystals such as beta barium oxide (BBO) \cite{Steinlechner2013,Steinlechner2014,Steinlechner2012a, Bedington2015c, Durak,Yin2017}.


A PPKTP-based design \cite{Steinlechner2012a} was identified by the European Space Agency for the original Space-QUEST proposal \cite{PerdiguesArmengol2008} and 
for the double-downlink demonstrations the Micius satellite uses a PPKTP source in a Sagnac interferometer arrangement, shown if Fig.~\ref{fig:sources} B \cite{Yin2017}. Sagnac-loop arrangements are auto-compensating allowing the spectral characteristics of the source to be tuned without requiring specifically modified birefringent walk-off compensation crystals \cite{Steinlechner2016}. The Micius source generates entangled photon pairs of $\sim$\SI{810}{nm} at a rate of $\sim$5.9 million per second under a pump power of $\sim$\SI{30}{mW}. It uses a thick titanium alloy baseplate and Invar mounts for thermal and mechanical stability. It also has piezo adjustable steering mirrors and an on-satellite BB84 receiver to allow it to be realigned in orbit \cite{Yin2017}.

Although BBO has a smaller $\chi^{(2)}$ non-linearity compared to PPKTP, its optical properties are much more temperature tolerant.
Furthermore, single-domain crystals are available in very large apertures (typically several millimetres), simplifying optical alignment, whereas periodically poled materials are restricted to a small aperture not exceeding \SI{2}{\mm} due to the difficulty in maintaining regular poling across the crystal.
Altogether, this makes BBO and other similar single-domain crystals very attractive raw material for the design of sources of entangled photon pairs \cite{Kwiat1995, Kwiat2000, Trojek2008}.

Polarization-correlated, photon-pair sources based on BBO have been demonstrated in orbit on board the 2U Galassia nanosatellite \cite{Tang2016}. The source design is also extremely robust, having survived a dramatic launch vehicle explosion \cite{Tang2015a}. The correlated source design is being extended and enhanced to generate polarization-entangled photon pairs \cite{Bedington2015c,Durak} for demonstration in future nanosatellite missions \cite{Bedington2015} (see Fig.~\ref{fig:sources}c).

The pump light from laser diodes for SPDC sources can easily reach \SI{40}{mW} when stabilized with external cavities \cite{ONDAX} and even exceed \SI{100}{mW} when operated in free-running mode. 
For QKD links using SPDC sources the limiting factor is therefore not the photon generation, but the ability of the single photon detectors to distinguish between the photons arriving with small timing separations, as will be discussed in section~\ref{sec:detectors}.

\subsection{Optical links}
\label{sec:opt}
The optical links use telescope-like optics at the photon source transmitter and at the receiver to beam the photons between satellite and ground station in the same manner as classical laser communications links.
The links are where the largest losses occur and thus have the biggest impact on the quantum bit error rate (QBER), which is the relevant signal-to-noise ratio in QKD and which will be discussed in more detail in section~\ref{sec:noise}. Simulated losses for example optical link scenarios are shown in Table~\ref{tab:losses} which is adapted from Ref. \cite{Aspelmeyer2003} and does not include detector losses.

\begin{table}
\centering
\caption{\label{tab:losses}Simulated link attenuation for \SI{800}{nm} photons in various link scenarios \cite{Aspelmeyer2003}}
\begin{tabular}{ | p{2.2cm} | p{2.1cm} | p{2.0cm} | p{2.1cm} |}
\hline
  &\textbf{ \SI{1}{m} ground receiver} & \textbf{ \SI{30}{cm} LEO receiver }& \textbf{ \SI{30}{cm} GEO receiver} \\
\hline
\textbf{\SI{1}{m} ground transmitter} &               & \SI{27.4}{dB}   (\SI{500}{km})  & \SI{64.5}{dB}     (\SI[group-separator = {,}]{36000}{km}) \\
\hline
\textbf{\SI{30}{cm}  LEO transmitter} & \SI{6.4}{dB}     (\SI{500}{km}) & \SI{28.5}{dB}     (\SI[group-separator = {,}]{2000}{km})    & \SI{52.9}{dB}    (\SI[group-separator = {,}]{35000}{km})   \\
\hline
\textbf{\SI{30}{cm}  GEO transmitter} & \SI{43.6}{dB}  (\SI[group-separator = {,}]{36000}{km})       & \SI{52.9}{dB}     (\SI[group-separator = {,}]{35000}{km})  & \SI{53.9}{dB}  (\SI[group-separator = {,}]{40000}{km})  \\
\hline
\end{tabular}
\end{table}
%

In the downlink configuration, optical loss is essentially dominated by diffraction, i.e. broadening of the beam, which increases as the square of the link length. In the uplink configuration, atmospheric turbulence, which is most significant in the \SI{20}{km} immediately above the surface of the Earth, has a much larger effect, adding over \SI{20}{dB} of loss for the example in Table~\ref{tab:losses}. This extra
loss in uplink may mitigate many of the perceived conveniences gained from locating the quantum light source on the ground. However, since this atmospheric turbulence is fluctuating, the use of signal-to-noise ratio filters to discard data when noise levels are highest, can lead to lower QBER and thus longer private keys \cite{Capraro2012,Usenko}. This is because if the full data set were used, the QBER value would be higher and many more of the photons would be required for error correction and privacy amplification. Similar schemes have also been discussed for CV-QKD \cite{Elser2012}.

The following subsections discuss the origins of these optical losses and the fundamental engineering challenges to address when establishing optical links. These are principally: coping with the losses/depolarizing effects in the optical path, and ensuring the telescopes are pointing precisely at each other.

\subsubsection{\textbf{Optics considerations}}
\label{sec:trans}
\begin{figure}[h!]
\centering
\includegraphics[width=\columnwidth]{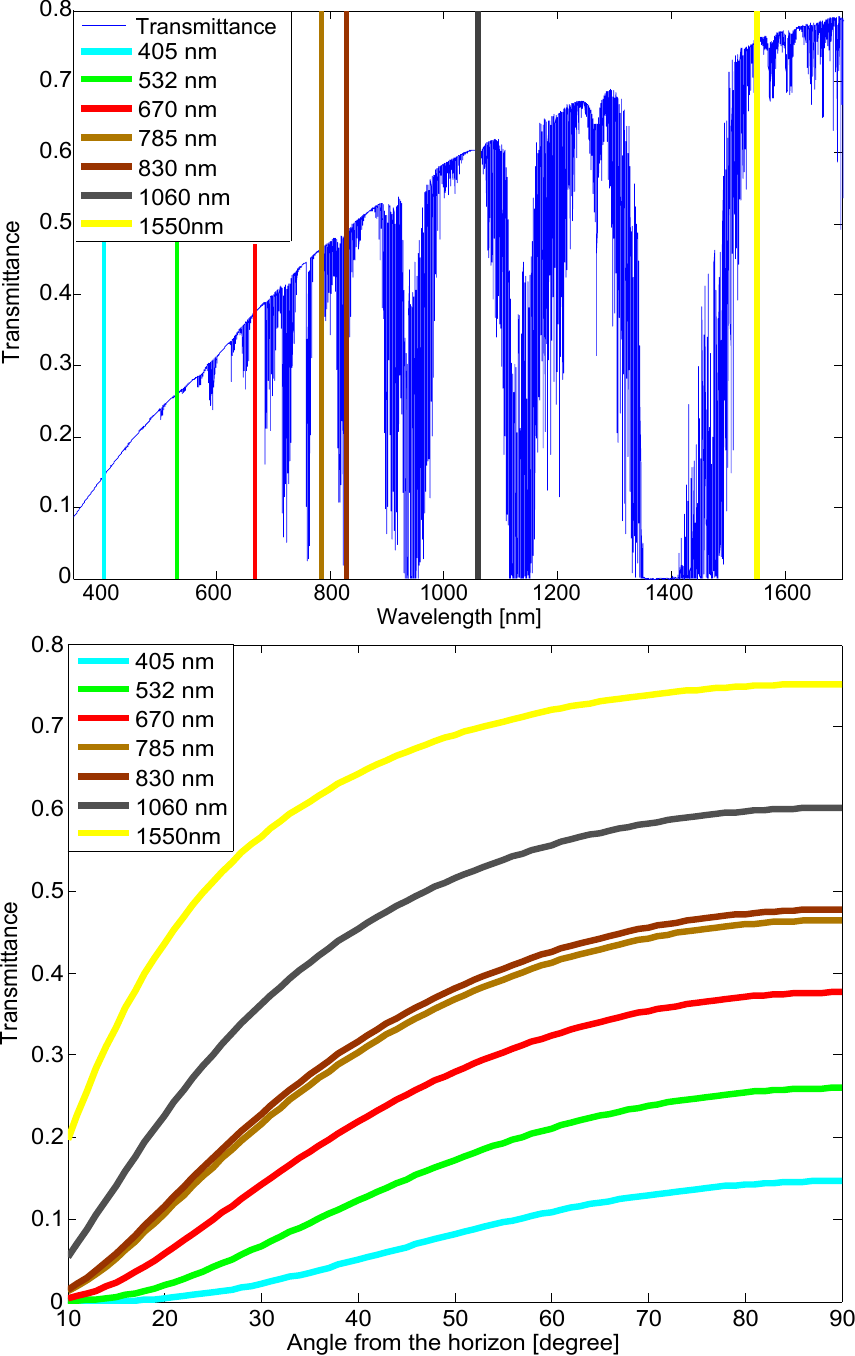}
\caption{\label{fig:atmosphere} Simulated transmittance of the atmosphere (adapted from \cite{Bourgoin2012}). Top --  at zenith in a typical rural location with overlaid coloured lines showing the wavelengths of commercially available laser systems. Bottom --  as a function of the pointing angle above the horizon for the various lasers.}
\end{figure}

Diffraction losses are dependent on the telescope design and beam spatial mode and  increase as the inverse square of the wavelength ($\lambda$). Conversely, atmospheric absorption losses tend to decrease with increasing wavelength, though there are also large spectral bands in which the atmosphere is almost entirely opaque.  Losses due to atmospheric turbulence also decrease slightly as wavelengths increase. 
Favourable spectral bands, where relevant laser systems also exist, can be seen in Fig.~\ref{fig:atmosphere} and include \SIrange{665}{685}{nm}, \SIrange{775}{785}{nm}, \SIrange{1000}{1070}{nm} and \SIrange{1540}{1680}{nm} \cite{Bourgoin2012}.

Of the methods discussed in section~\ref{sec:concepts}, a single downlink is the most practical, and in these cases entanglement-based QKD has been shown to be more tolerant to loss than prepare-and-measure schemes due to the intrinsic timing correlation between photon-pairs generated in the SPDC process \cite{Scheidl2009}.

A trade-off when designing  the telescope optics is the choice of either transmissive or reflective telescopes. Reflective mirrors can be larger, although for polarization-based QKD care must be taken to prevent large depolarization effects. Secondary mirrors of reflector telescopes are often placed within the path of the beam so that it blocks part of the primary mirror and affects the diffraction spreading, although the loss is less than \SI{1}{\dB} for a secondary mirror diameter up to 25\% the ratio of the primary mirror diameter ($D$) \cite{Bourgoin2012}.
Beam divergence scales with $\lambda/D$, so larger apertures produce smaller beam divergences with lower free-space losses, although because of atmospheric turbulence, increasing the diameter of ground-based transmitters above a few tens of centimetres makes little improvement on the losses. Indicative examples of ground-based transmitting telescopes in studies start from \SI{25}{cm} in diameter, LEO-based transmitting telescopes from \SI{9}{\cm} and GEO-based from \SI{13.5}{\cm} \cite{Elser2012,Bourgoin2012,Oi2017}. The Micius quantum satellite uses transmitters of 18 and \SI{30}{cm} \cite{Yin2017}.


The success of the Micius satellite has confirmed the scientific consensus \cite{Toyoshima2009,Vallone2015a} that the atmosphere does not degrade polarization states. Additional studies have also suggested this to be true for the time-bin degree of freedom \cite{Vallone2016}.

The Micius satellite collects its entangled photons into single mode fibres. These are sensitive to vibrations and after launch introduced random polarization rotations which needed to be compensated using motorized waveplates during in-orbit commissioning and every few weeks thereafter. The satellite's transmitter also has additional polarization correcting elements to compensate for the spacecraft attitude drift. Other polarization corrections are usually performed at the receiver and are discussed in the receiver section (\ref{sec:rec}) below. Time-bin entanglement-based QKD would require additional Doppler corrections \cite{Zeitler2016}.

\subsubsection{\textbf{Pointing, acquisition and tracking (PAT)}}
\label{sec:pat}

Establishing the optical link is usually performed in several stages. Satellite orbit determination data such as radar tracking, GPS and star tracker measurements can be exchanged between satellite and ground via radio frequency (RF) links to provide a coarse level of mechanical pointing towards each other. Laser beacons on the ground and on the satellite can then be used as a target to enable a finer level of mechanical pointing. The finest level of pointing can then be achieved through optical beam-steering systems, which are essentially correcting for atmospheric turbulence. 

Full deformable mirror adaptive optics systems could be implemented, but they are typically not designed to preserve polarization, and since the beam is small, higher order wavefront perturbations would not make a large impact \cite{Oi2017}. For larger spacecraft, coarse level pointing is usually achieved by mounting the telescopes on two-axis gimbal or turntable stages, whereas for nanosatellites coarse control is usually achieved by reorienting the entire satellite. 
Such stages are used together with piezo fast steering mirrors on the Micius satellite to achieve a final pointing of around \SI{0.6}{\micro \radian} in space \cite{Yin2017}. CQuCom (CubeSat Quantum Communications Mission) proposes a \SI{3}{\micro \radian} pointing capability \cite{Oi2017}, which would push the boundaries of current CubeSat capabilities.

Relative to the incoming laser beacons, outgoing optical signals must be pointed at an angle to compensate for the time delay and chromatic dispersion.
For typical ground-to-LEO QKD uplink proposals, atmospheric turbulence means the transmitter pointing accuracy has less of a consequence than for LEO-to-ground downlinks. A \SI{2}{\micro rad} rms error in the pointing of a \SI{20}{cm} downlink transmitter would introduce \SI{4}{dB} of loss compared to less than \SI{1}{dB} for a \SI{20}{cm} uplink transmitter \cite{Bourgoin2012}. Jitter and imperfections in the tracking systems should be minimized so their contributions to beam broadening are much less than those caused by diffraction and atmospheric turbulence. The receiver system need only point to an accuracy within its field of view, e.g. \SI{50}{\micro rad} \cite{Bourgoin2012}.




\subsubsection{\textbf{Optical receivers}}
\label{sec:rec}

Optically, the receiver telescope can be identical to the transmitter telescope, discussed in section~\ref{sec:trans}.
For uplink configurations, the final key rate is strongly driven by the diameter of the space-based receiver telescope \cite{Bourgoin2012}, but flying large telescopes in space is complex and costly.
For uplink QKD experiments to the International Space Station (ISS), a \SI{14.3}{\cm} camera lens mounted on an ESA NightPod tracking stage has been proposed as a QKD receiver \cite{Scheidl2013}. 
The Canadian NanoQEY uplink nanosatellite proposal uses a similar-sized \SI{15}{\cm} receiver while their larger (microsatellite) uplink proposal, QEYSSAT, opts for a \SI{40}{\cm} one. The QEYSAT receiver would be similar to the quad detector receiver shown in Fig.~\ref{fig:sources}a, so it can potentially be made capable of performing both BB84 and BBM92 protocols, to allow for a variety of QKD sources to be demonstrated with it \cite{Jennewein2014d}.

Ground-based receivers can be made larger more easily than their space-based counterparts and have a large impact on key rates for downlink configurations \cite{Bourgoin2012}.
In retroreflector experiments, the \SI{1.5}{m} Matera Laser Ranging Observatory (MLRO) was used to receive the qubits reflected from satellites from \SIrange{1000}{7000}{\km} \cite{Vallone2015, Vallone2016}. 
MLRO was also baselined as the receiver for the CQuCom mission \cite{Oi2017}, while the Micius satellite has so far been using ground-based receivers of \SI{1}{m}, \SI{1.2}{m} and \SI{1.8}{m} \cite{Yin2017,Sheng2017}.
For commercial usage though, smaller diameters are considered more attractive (e.g. \SIrange{25}{50}{\cm}), in part because they can be made more mobile, except for GEO communications where receivers can have fixed pointing and thus are easier to make larger (e.g. \SI{2}{m}) \cite{Elser2012}. While free space QKD demonstrations are sometimes performed at high altitude sites to reduce atmospheric turbulence, these locations are unlikely to be attractive for useful QKD networks.
For commercial usage though, smaller diameters are considered more attractive (e.g. \SIrange{25}{50}{\cm}), in part because they can be made more mobile, except for GEO communications where receivers can have fixed pointing and thus are easier to make larger (e.g. \SI{2}{m}) \cite{Elser2012}. While free space QKD demonstrations are sometimes performed at high altitude sites to reduce atmospheric turbulence, these locations are unlikely to be attractive for useful QKD networks.

For polarization-based QKD schemes, the receiver telescope is typically coupled to an analyser like that in Fig.~\ref{fig:sources}a or, in the case of the Micius entanglement distribution demonstration, a Pockels cell connected to a random number generator \cite{Yin2017}. For the analysers to be effective, systems must be in place to allow dynamic polarization variations -- most particularly the relative roll orientation of the satellite and ground station -- to be understood and compensated, so that the reference frame of the polarization bases is preserved. This is a complex task for non-GEO orbits as the satellites are moving across the sky throughout the operation, but it can be achieved  without the requirement for a feedback loop between space and ground \cite{Bonato2006, Toyoshima2011, Wang2014, Carrasco-Casado2016}. In all tests, atmospheric depolarization has been found to be minimal. 

\subsection{Detection and key generation}
\label{sec:detect}

\begin{figure*}[t!]
\centering
\includegraphics[width=\textwidth]{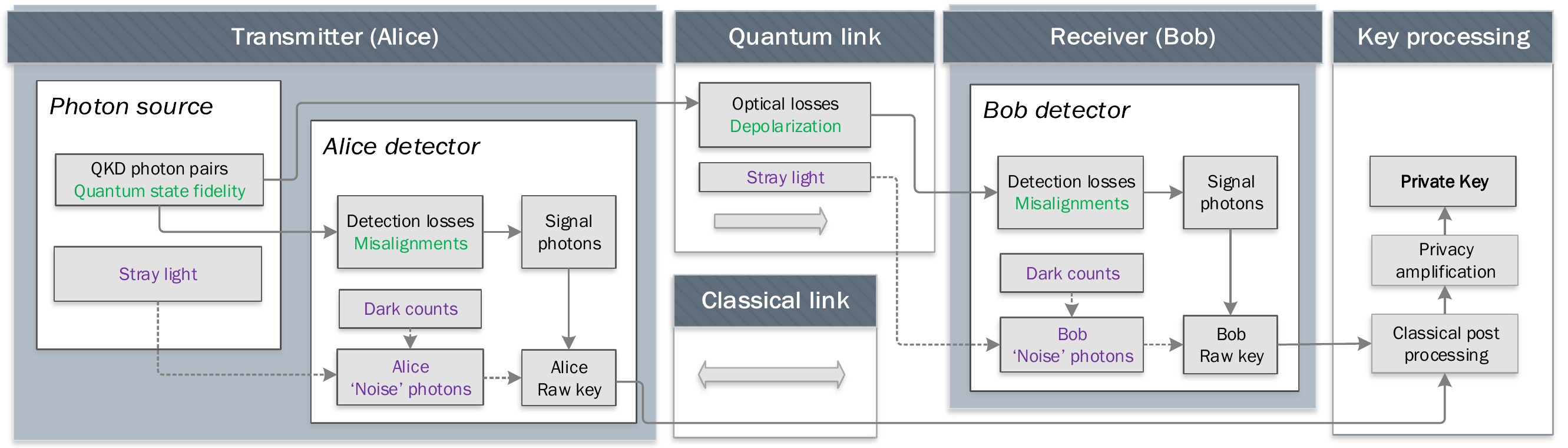}
\caption{\label{fig:qber-block} Noise contributions to entanglement-based QKD between Alice (transmitter) and Bob (receiver) in a scenario such as Fig.~\ref{fig:protocol} frame (a). Effects determining the fidelity of the quantum states (visibility) are shown in green. Sources of unwanted photons are shown in purple with dotted connector lines. A WCP BB84 configuration would look similar, except Alice would have a random number generator selecting the polarization bases for her photon source and would not have a detector.}
\end{figure*}

The optical signal arriving at the receiving detectors is composed of both QKD photons and stray light. The detector output is also mixed with technical noise (e.g. `dark-counts' from GM-APDs) and is considered as a raw key that must be processed to produce a private encryption key (see Fig.~\ref{fig:qber-block}).

\subsubsection{\textbf{Detectors}}
\label{sec:detectors}

In DV-QKD schemes, the individual photons are registered by photo multiplier tubes (PMTs) or increasingly by GM-APDs -- also known as single-photon avalanche diodes (SPADs) -- which are advantageous due to their smaller power requirement and physical footprint. 


As has been mentioned in section~\ref{sec:sources}, the detector time resolution capability is typically the system performance bottleneck rather than the source brightness. Silicon-based GM-APDs have a typical timing jitter of \SI{0.5}{ns}; a reduced timing jitter can be achieved by sacrificing the avalanche volume of the photodiode, which means that the detection efficiency is also reduced.

Detectors also have a recovery time (dead time) between detections, for APDs this can be e.g. \SI{26}{ns} \cite{Stipcevic2013}. CMOS arrays of GM-APDs offer the potential to reduce this to several tens of picoseconds \cite{Oi2017}.

Space-based GM-APDs must be appropriately shielded from radiation damage if several years of operation are required.
The detection efficiency of GM-APDs is typically on the order of 50\% for silicon-based devices, which can detect visible or near-infrared light. 
GM-APDs for IR telecom wavelengths also exist, although they are much noisier and have a lower detection efficiency -- around 20\% -- due to the different materials needed to detect long-wavelength photons. 

\begin{figure*}[t!]
\centering
\includegraphics[width=\textwidth]{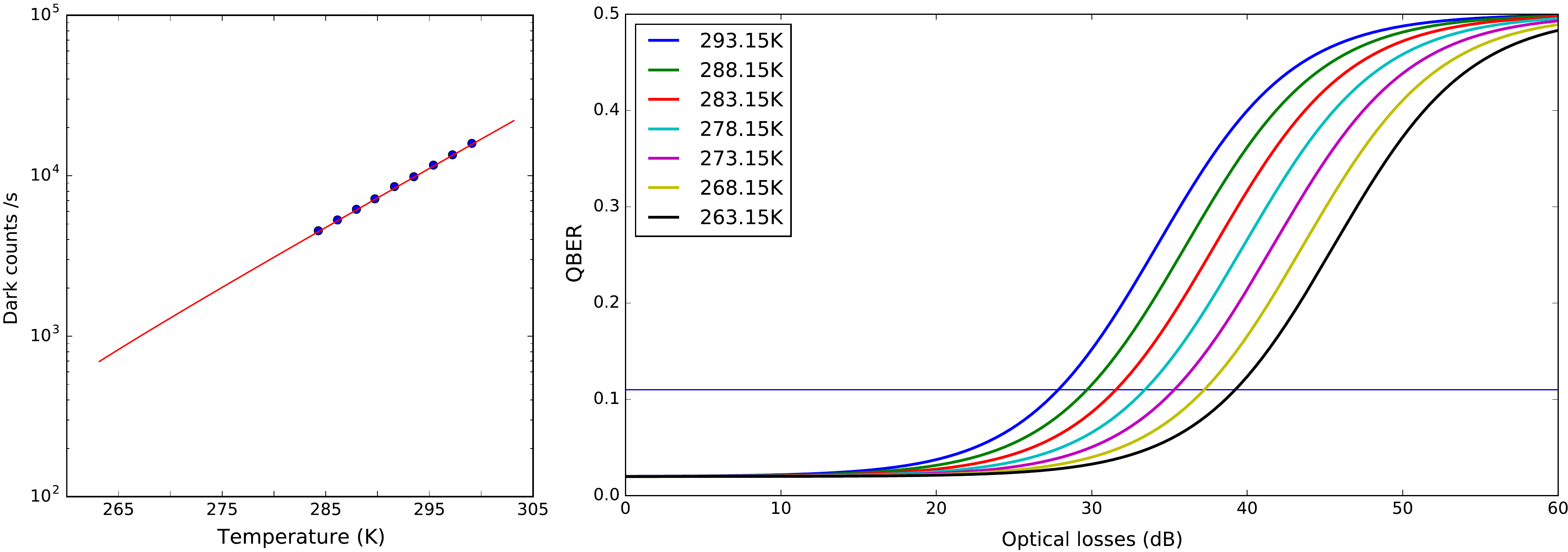}
\caption{\label{fig:temp} Impact of noise and losses on an example BBM92 system. \textbf{Left} -- dark counts measured in a typical SAP500 GM-APD as a function of temperature. Blue dots are measurements, red line is fitted exponential, $ae^{(x-273.15)b}+c$ where $a=1790, b=0.08, c=-81$. 
\textbf{Right} --  QBER as a function of optical losses (e.g. diffraction, atmosphere, etc.) and technical noise in the SAP500 detector at different temperatures. Secure QKD is only possible when QBER is below 11\% (dotted line).}
\end{figure*}

The GM-APDs have noise from dark-counts which have an exponential dependence on temperature as demonstrated in Fig.~\ref{fig:temp} (left), and which can be a significant contributor to QBER in the satellite environment.
Fig. \ref{fig:temp} (right) shows that a simulated QKD link can tolerate an additional \SI{0.5}{dB} of loss for every degree drop in temperature, making it highly attractive to provide cooling to the detectors.

In addition to PMTs and GM-APDs, ground-based receivers equipped with cryostats do have the option of operating state-of-the-art single-photon detectors which use superconducting technologies and can have detection efficiencies over 90\%. 
These detectors can have extremely low technical noise, but must be appropriately shielded from thermal blackbody emission.
While  superconducting detectors need to be cooled to \SI{4}{K} or less, for optimal performance GM-APDs need only to be cooled to about \SI{250}{K}.
Unless the ultra-low temperature requirements are overcome, superconducting detectors are unlikely to be an attractive option for use in space \cite{Miki2010}.

Thermal management of any space-based device can be challenging. In-orbit temperatures can vary by tens of degrees, electrical power is limited, and, for the most part, heat can only be lost radiatively. 
One approach for the APDs may be to cool them by connecting them passively to radiators on the satellite, and while excursions in temperature may occur, stable operation can still be achieved by monitoring the pulse height of the detector output to actively control the GM-APD bias voltage \cite{Tan2013}. 
Technical noise in the detectors steadily increases as ionizing space radiation damages the silicon with extended time in orbit \cite{Tan2015}. Efforts are underway to investigate effective shielding techniques, as well as in-orbit annealing methods to attempt to reverse this effect \cite{capsat, Anisimova2017, Lim2017a}.


\subsubsection{\textbf{Quantum Bit Error Rate (QBER)}} \label{sec:noise}

QBER is the percentage of the sifted raw keys that is erroneous, i.e. that does not match between Alice and Bob. 
 As the rate of the true signal falls, the accidental coincidences from stray light and other technical noise in the detection and measurement apparatus make a larger contribution to the raw keys and the QBER rises. Once QBER exceeds 11\%, the QKD protocols based on BB84 -- which are the ones considered by the vast majority of satellite projects -- will not be able to generate any private key.

For DV-QKD protocols, noise appears as accidental correlations between the detectors of Alice and Bob.
Accidental correlations are well-approximated by the expression: $S_1 \times S_2 \times \tau$, where $S_1$ and $S_2$ are the rate of events at the individual detectors, and $\tau$ is the coincidence-time window. In systems where either detector can fire first, an additional factor of 2 is needed \cite{Janossy1944}. This expression only works when the detectors are in the linear regime.
In scenarios where the detectors are saturated, e.g. by stray light, the recovery behaviour of the detectors must be taken into account \cite{Grieve2015a}.

The effect of link loss on QBER for a model BBM92 system is shown in Fig. \ref{fig:temp} (right) where the QBER is plotted against link loss for a range of GM-APD temperatures, with the corresponding dark counts shown in Fig.~\ref{fig:temp} (left). In this simulated system, the intrinsic QBER is assumed to be 1.5\% at full transmission, the source produces photon pairs at \SI{1}{Mcps} and the coincidence window is \SI{2}{ns}. 

Although it is not specifically included in the simulation in Fig. \ref{fig:temp}, another significant noise source is stray light: background photons being detected. In general, stray light can be minimized in optical systems with extensive baffling and optical blacking, however even after extensive filtering \cite{Peloso2009, Hughes2002a}, DV-QKD is likely to be a night-time activity for the near future \cite{Er-long2005a,Tomaello2011}. A possible, but challenging, solution to this is to perform DV-QKD at other wavelengths with alternative detectors \cite{Sheng2016}, or -- if suitable light sources can be found --  at the wavelengths of the Fraunhofer lines, the narrow absorption lines in the output spectra of the Sun \cite{Benton2010}. 
In this respect, CV-QKD has the advantage because the optical systems used have a sufficiently small spectral bandwidth that allows much of the background to be filtered, enabling daytime operations \cite{Gunthner2016b}.

\subsubsection{\textbf{Establishing keys}}

For QKD to be carried out, a number of communication tasks need to be performed between the key sharing parties, such as  basis reconciliation, clock synchronization and time tagging of photons, the latter being the most significant in data bandwidth. 

For QKD based on SPDC sources the intrinsic timing correlation of entangled photon pairs \cite{burnham1970} is in the femtosecond regime \cite{Hong1987}, and can be exploited to perform asynchronous pair detection over long-range \cite{Ho2009}. Prepare-and-measure schemes require alternative approaches \cite{Bienfang2004,Bourgoin2015a}; the Micius satellite achieves this through \SI{10}{kHz} pulses in its \SI{532}{nm}, downward-pointing tracking beacon \cite{Sheng2017}. 
Laser ranging to passive retroreflectors on the satellite can also be used to provide centimetre level distance knowledge \cite{Oi2017}.

\begin{figure}[h]
\includegraphics[width=\columnwidth]{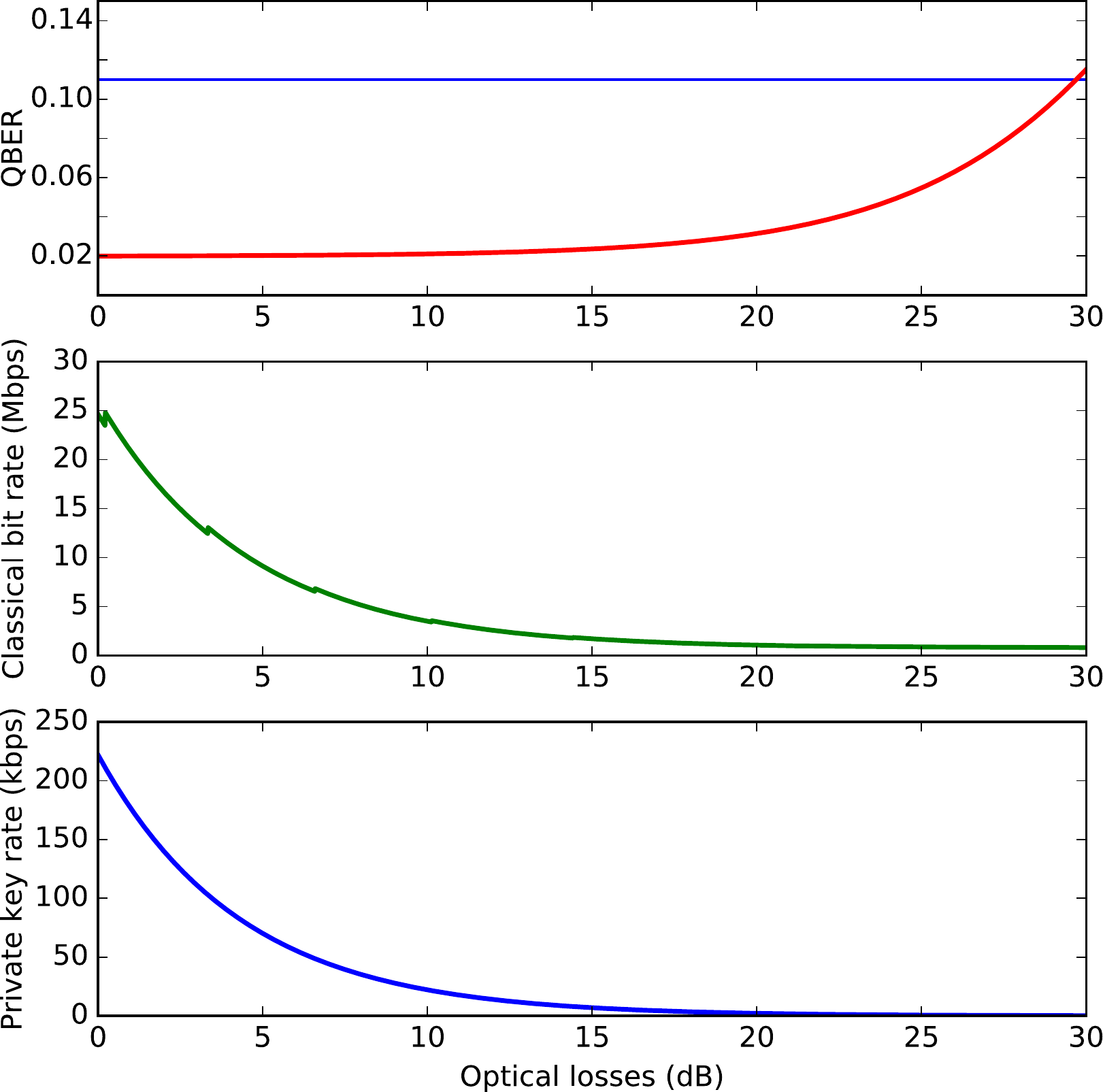}
\caption{\label{fig:bitrates} Private key rate achievable as a function of optical losses, and the corresponding classical communication overhead that would be required for continuous steady state key production, for the simulated BBM92 link in the previous figure (\SI{288.15}{K}). The highest QBER with which QKD can be performed securely here is 0.11 (horizontal blue line).}
\end{figure}

For continuous key production in the simulated BBM92 system used for Fig.~\ref{fig:temp}, with a detector temperature of \SI{288.15}{K}, the classical communications link requires a baseline bit rate of a few Mbps which would increase to a few tens of Mbps as the QBER decreases allowing for a few 100s  kbps of private key to be established, as shown in Fig.~\ref{fig:bitrates}.  

For the specific example of the losses in the LEO-to-ground configuration in Table~\ref{tab:losses} this would correspond to about \SI{55}{kbps} of private key requiring about \SI{7.5}{Mbps} classical bit rate for steady state operations.
For space-based QKD this authenticated classical link could be established over radio links, or since an optical link is already established it could be implemented using classical laser communications. The Micius satellite opts for radio with \SI{1}{Mbps} uplink and \SI{4}{Mbps} downlink \cite{Sheng2017}.

However, for QKD there is no requirement that the classical communications happen at the same time as the quantum communications, i.e. for non-steady state key production, the classical communications could be stored and transmitted at a later time and the time stamps synchronized, bases matched and keys generated in post processing. For example, after passing over an optical ground station and transmitting or receiving qubits, a LEO satellite could then perform the classical comms at a slower rate, e.g. via a GEO communications relay during the remainder of the orbit.

The proposed NanoQEY uplink QKD nanosatellite is projected to create \SI{10}{kbit} of secure key per month with its \SI{15}{\cm} receiver, paired with a \SI{50}{\cm} telescope on the ground \cite{Jennewein2014}. The microsatellite QEYSSAT uplink proposal however, suggests that a \SI{40}{\cm} receiver can achieve $\sim$\SI{100}{kbit} of secure key per pass \cite{Jennewein2014d}. In contrast, the Micius downlink satellite, with a \SI{30}{cm} transmitter in space and a \SI{1}{m} receiver on ground, is achieving $>$\SI{300}{kbit} of secure key per pass. For future QKD service satellites it has been estimated that with a \SI{13.5}{\cm} transmitter at LEO, the monetary costs of QKD for \SI{50}{\cm} and \SI{25}{\cm} receivers could be \EUR{77} per Mbit and \EUR{312} per Mbit respectively while a service from GEO to a \SI{2}{m} receiver would be \EUR{615} per Mbit \cite{Elser2012}.

To operate as a trusted node within a QKD network additional elements such as key management, and control and accounting subsystems are required \cite{Elser2012}. As with all practical implementations of QKD, special considerations must be paid to the side-channel attacks that might be possible with real-world systems \cite{Scarani2014}.


\section{Conclusions}

Satellite QKD has overcome the range limit of ground-based links and, is being used to enable global coverage \cite{Sheng2017}.
The technical challenges to achieving a global network remain daunting, but progress is being made in overcoming them.

Table~\ref{table: Summary} shows a summary of notable satellite QKD initiatives that have been reported in recent years.

\begin{table*}
\begin{center}
\caption{\label{table: Summary} A summary of satellite QKD enabling initiatives}
		\setlength\extrarowheight{1.15pt}
    \begin{tabular}{ | p{2.4cm} | p{6.1cm} | p{2.8cm} | p{6cm} |}
    \hline
    \textbf{Initiative} & \textbf{Goal} & \textbf{Vehicle} & \textbf{Status/Results} \\  \hline
		   QUESS \cite{Pan2014}& LEO-to-ground trusted-node satellite QKD, uplink quantum teleportation and double-downlink entanglement distribution.
     & Micius \SI{631}{kg} satellite. &Entanglement distribution of \SI{1203}{km} \cite{Yin2017}, teleportation up to \SI{1400}{km} \cite{Ren2017} and BB84 QKD up to \SI{1200}{km} with QBER $\sim1\%$ \& sifted key \SI{14}{kbps} \cite{Sheng2017}.\\ \hline 
   Toyoshima et al. \cite{toyoshima2009polarization} & LEO-to-ground polarization measurement & OICETS \SI{570}{kg} satellite.& Polarization preserved within system rms error of \SI{28}{mrad}. \\ \hline
    Takenaka et al. \cite{Carrasco-Casado2016, Takenaka2017} & LEO-to ground polarization and quantum limited measurements from a small optical transponder (SOTA). & SOCRATES \SI{48}{kg} microsatellite. & Effectively no depolarization was observed (100\% Degree-of-polarization) and QBER of $<5$\%\\ \hline
    G\"{u}nthner et al. \cite{Gunthner2016b} & GEO-to-ground test of quantum state used in coherent communication. & Alphasat I-XL \SI{6649}{kg} satellite. & Quantum-limited states arrive on the ground after transmission from satellite. \\ \hline
 Vallone et al. \cite{Vallone2015} & Test of polarization state for weak coherent pulses using retro-reflectors on LEO satellites. & Jason-2 \SI{510}{kg}, Larets 21 kg and Starlette/Stella \SI{48}{kg} satellites. & Average QBER of 6.5\% achieved.\\ \hline    
    Yin et al. \cite{Yin2013a} & Test of polarization state for weak coherent pulses using retro-reflectors in a LEO satellite. & CHAMP \SI{500}{kg} satellite. & Signal to noise ratio of 16:1 observed for polarization measurements. \\ \hline
       Dequal et al. \cite{Dequal2016} & Test of weak coherent pulse transmission from retro-reflectors on a MEO satellite. & LAGEOS-2 \SI{411}{kg} satellite.& Peak signal-to-noise ratio of 1.5 with 3 counts per second.\\ \hline   	
    Tang et al. \cite{Tang2016} & In-orbit observation of polarization correlations from a photon-pair source on a nano-satellite. & Galassia 2 kg 2U CubeSat. & 97\% contrast in polarization correlation measurements. Pathfinder for SpooQySats (below).\\ \hline
   			Nauerth et al. \cite{nauerth2013air} & QKD between the ground and a aircraft moving at similar angular velocities to a LEO satellite.  & Dornier 228 utility aircraft. &  Sifted key rate of \SI{145}{bps}, QBER of 4.8\% from range of \SI{20}{km} at angular speed of \SI{4}{mrad} per second.\\    \hline
   			Bourgoin et al. \cite{bourgoin2015free}	& QKD with a moving receiver similar to the angular speed of satellite at \SI{600}{km} altitude. & Pick-up truck.  &  Key rate of \SI{40}{bps} with QBER of 6.5\% to 8\% with receiver at a range of \SI{650}{m} moving at angular speed of \SI{13}{mrad} per second.\\    \hline	
		Wang et al. \cite{wang2013direct} & Verification of pointing, acquisition and tracking.  & Hot-air balloon.  &  Key rate of \SI{48}{bps} and QBER of $\sim4\%$ over a range of \SI{96}{km}. \\    \hline
 
		  SpooQySats \cite{Bedington2016} & Demonstrate polarization-entangled photon-pair sources in space & 3U CubeSats & Funded mission. Launches planned from 2018.\\    \hline
	
		QEYSSat \cite{Jennewein2014d} & Trusted-node receiver for uplink QKD. & Microsatellite & Funded mission.\\ \hline
				    CAPSat \cite{capsat} & Laser annealing of radiation-damaged APDs  & 3U CubeSat  & Funded mission.\\    \hline
						NanoBob \cite{nanobob} & Trusted-node receiver for uplink QKD & CubeSat  & Proposal.\\    \hline
    SpaceQUEST (2008) \cite{PerdiguesArmengol2008} & Double LEO-to-ground downlinks QKD using polarization-entangled photon-pairs. & International Space Station & Proposal (since updated as a mission exclusively investigating decoherence due to gravity \cite{Joshi2017})
		\\ \hline
    Scheidl et al. \cite{Scheidl2013} & Entanglement-based QKD and Bell tests, ground-to-LEO. & International Space Station & Proposal.\\ \hline
		    
			 
     NanoQEY \cite{Jennewein2014} & QKD and Bell tests ground-to-LEO with a trusted-node satellite. & Based on NEMO nanosatellite bus, \SI{16}{kg}. & Proposal.\\ \hline
     Zeitler et al. \cite{Zeitler2016} & Superdense teleportation, LEO-to-ground & International Space Station & Proposal. \\ \hline
         QuCHAP-IDQuantique & Establish QKD networks based on high altitude platforms. & High-altitude platform. &  Proposal.\\    \hline
		    CQuCom  \cite{Oi2017} & LEO-to-ground QKD downlinks & 6U CubeSat  &  Proposal.\\    \hline
    \end{tabular}
\end{center}
\end{table*}
The top eleven rows show completed or ongoing experiments and missions.  Besides photon sources and tracking systems \cite{Durak,Tang2016,Vest2015}, much important work has been made to confirm that quantum states encoded with polarization \cite{Vallone2015,Wang2014} and time-bin \cite{Vallone2016}  experience negligible decoherence in space-ground optical links. These are showing that the standard QKD protocols, already proven in ground tests, also work from space, possibly even from geostationary orbits \cite{Gunthner2016b}. The most notable experiment thus far is QUESS (the Micius satellite) by the Chinese Academy of Sciences \cite{Pan2014}.


Work in developing light sources for DV-QKD is ongoing at various groups. For protocols relying on weak coherent pulses, there are now very compact designs using laser diodes and waveguide-based polarization rotators \cite{Vest2015}. Space capable polarization-entangled photon-pair sources to enable BBM92 or E91 protocols are also at various stages of development \cite{Durak,Steinlechner2016}. 

The performance of single photon detectors is an important consideration for DV-QKD. Currently, the limitation on QKD rates (in space or on the ground) is not given by the  brightness of the photon sources, but by the detection efficiency and timing jitter of the single photon detectors.
At the moment the only fast detectors with near-unit detection efficiency are based on superconducting technology and have requirements that make them costly to operate on satellites.
This is motivating research into GM-APDs to understand how to operate them more efficiently, and to prolong their lifetime in a space environment \cite{Tan2013, Lim2017a}.

Table~\ref{table: Summary} also lists the sizes of the satellites used. In recent years research on satellite QKD has become more attractive with the emergence of nanosatellite platforms, particularly the CubeSat. The standardized satellite buses provided by CubeSats allow for the rapid development of cost-effective space missions. 
In the first instance, these missions aim to be technological pathfinders aimed at raising the technology readiness level of sub-systems.
As the technology matures, we can expect that full QKD missions with nanosatellites, such as the NanoQEY \cite{Jennewein2014} and CQuComm \cite{Oi2017} proposals, will be realized.

The placement of quantum optics technologies on satellites also enables various fundamental physics experiments \cite{Rideout2012}, such as placing experimental bounds on the effect of gravity on quantum systems as they traverse a changing gravitational potential \cite{Ralph2014a, Joshi2017} . 

A technology related to QKD is the quantum random number generator \cite{Ma2015} used in the production of random bits that are not pre-existing. This technology already has applications in a number of cryptographic scenarios, relevant in both ground and space environments. The distribution of entanglement as part of a quantum internet \cite{Kimble2008,Pirandola2016} is another possibility---the capability to perform this is an important building block in a global network of distributed quantum computers. 

A quantum internet would not be a purely optical technology like QKD, but will require an interface to material systems that can act as quantum memory or processing units. Furthermore, in contrast to QKD, Doppler effects cannot be neglected in any attempted satellite information transfer because candidates for quantum storage can only interact with light fields with very narrow linewidths (in the kHz or MHz regime). 

Another potential area for growth is the placement of quantum sensors on satellites to aid in gravimetry or mineral exploration. First steps in operating quantum matter systems in space are being carried out by fundamental physics experiments on the ISS \cite{Soriano2014,Williams2015}, on sounding rockets \cite{Schkolnik2016} as well as with nanosatellites \cite{Oi2016}. 	

In summary, quantum satellite communications is paving the way for the deployment of other quantum technologies in space. The current pace of
development suggests that world-wide communications privacy can be maintained in the era of powerful quantum computers, and at the same
time it is bringing the concept of a global quantum internet closer to fruition.









\begin{acknowledgements}
The authors thank Tanvirul Islam and Christian Kurtsiefer for feedback and Tanvirul Islam for the QBER simulation and plots.\\

\textbf{Data Availability}\\
The code for the QBER simulation can be found here: https://github.com/CQT-Alex/s2qkd-sim \\

\textbf{Contributions} \\
All authors contributed to the paper equally.
RB led section~IV, JMA led sections~I-III, AL led section~V.\\
All authors then reviewed and contributed to each others sections substantially.\\


\textbf{Competing Interests} \\
The authors declare that they have no competing financial interests.\\

\textbf{Funding} \\
This work is partially supported by the National Research Foundation, Prime Minister's Office, Singapore (under the Research Centres of Excellence programme and through Award No. NRF-CRP12-2013-02) and by the Singapore Ministry of Education (partly through the Academic Research Fund Tier 3 MOE2012-T3-1-009).\\

\end{acknowledgements}

\end{document}